\documentclass[a4paper]{jpconf}
\usepackage{graphicx}
\usepackage{multirow}
\begin{document}
\title{Effects of Jets in the Flow Observables}

\author{Jun Takahashi, Rafael Derradi de Souza and David D. Chinellato}

\address{Instituto de F\'isica Gleb Wataghin, Universidade Estadual de Campinas, Rua S\'ergio Buarque de Holanda 777, 13083-859 S\~ao Paulo, Brazil}

\ead{rderradi@ifi.unicamp.br}

\begin{abstract}
The transverse momentum anisotropy of the particles produced in heavy ion collisions is one of the most important experimental observable to investigate the collective behavior of the systems created in such collisions. Recent studies show that the complex nature of the system evolution, such as initial condition fluctuations and jets, may lead to important effects in the flow coefficients and, therefore, to misinterpretation of the results obtained. In this study, we used simulated events produced with a hydrodynamic model which allows inhomogeneous initial condition combined with proton-proton collisions produced with the Pythia event generator to create a final set of particles to be analyzed with the usual experimental flow calculation techniques. Although this simplified approach is somehow unrealistic, since it does not include the interaction of the jet with the medium, our results have shown a good agreement of the behavior of the elliptic flow coefficient as a function of the transverse momentum up to 6 GeV/c for Au+Au collisions at 200 GeV. Although each model alone is not able to describe the full range, the combination of both sets of particles as seen by the flow calculation techniques may be the key to explain the behavior observed in experimental data.
\end{abstract}

\section{Introduction}
One of the main observables related to the collective behavior of the system created in relativistic heavy ion collisions is the transverse momentum anisotropy of the produced particles.
Ideally, one would measure the azimuthal distribution of particles with respect to the reaction plane directly, but this is not experimentally feasible and there is a need for techniques to compensate for this missing quantity based on azimuthal angle correlations of the particle momenta.
In this context, recent studies have demonstrated that the system evolution may contain interesting aspects such as initial condition inhomogeneities and jets of particles originating from hard parton-parton interactions.
These aspects could have important adverse effects on the flow coefficients and thus merit further investigation.

In this study, we combine two decorrelated components, one hydrodynamic-like and another jet-like part, such that the resulting event may be analyzed with common experimental calculation techniques, allowing us to explore any effects that may arise due to the inclusion of jets.

\section{NeXSPheRIO + Pythia}
\begin{figure}[h]
\begin{minipage}{15pc}
\includegraphics[width=14pc]{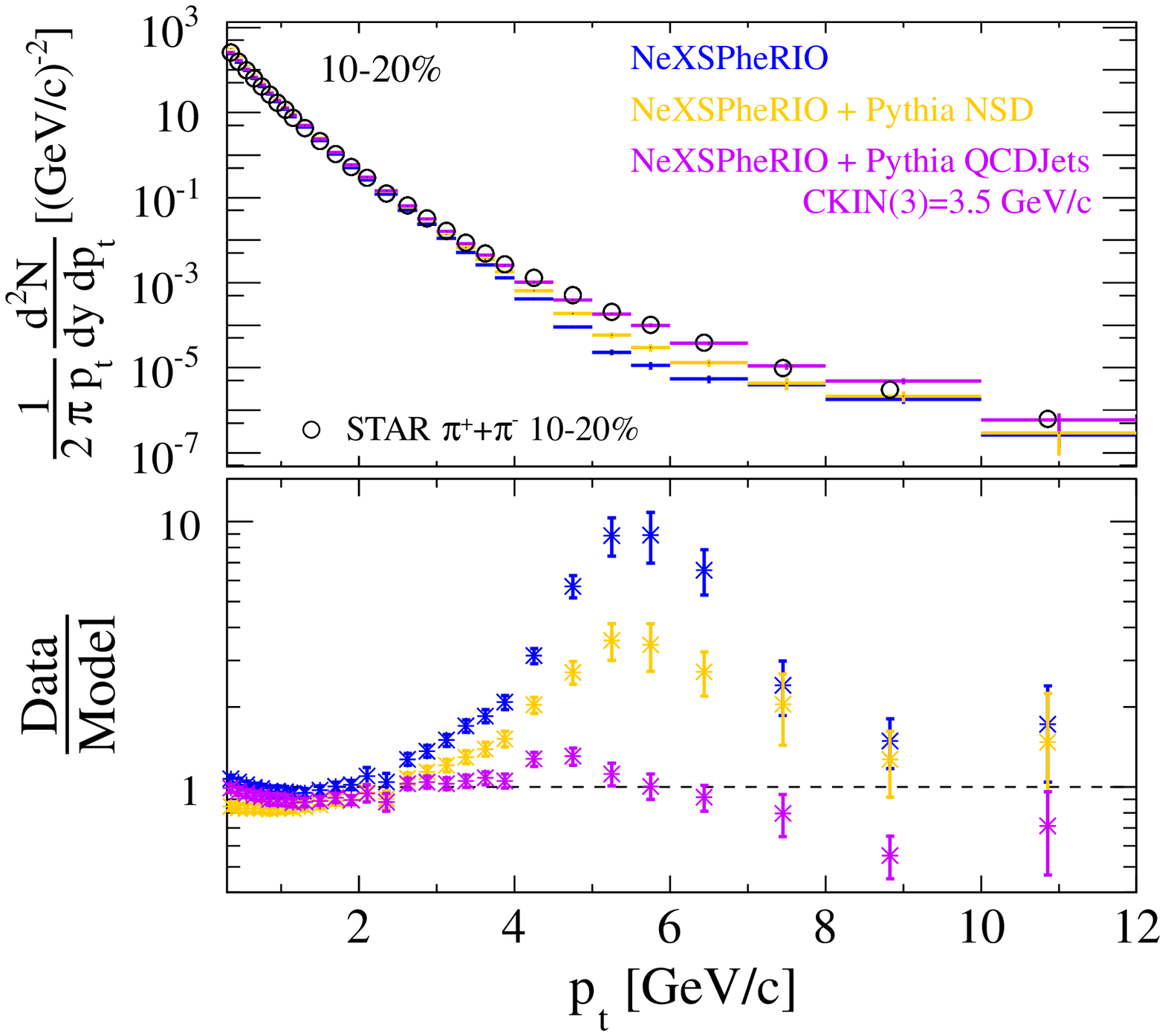}
\caption{\label{fig:InvYield_vs_pt}Invariant $p_t$ spectra of charged pions for NeXSPheRIO+Pythia model compared to experimental results from STAR \cite{Abelev:2006jr}.}
\end{minipage}\hspace{2pc}
\begin{minipage}{17pc}
\includegraphics[width=15pc]{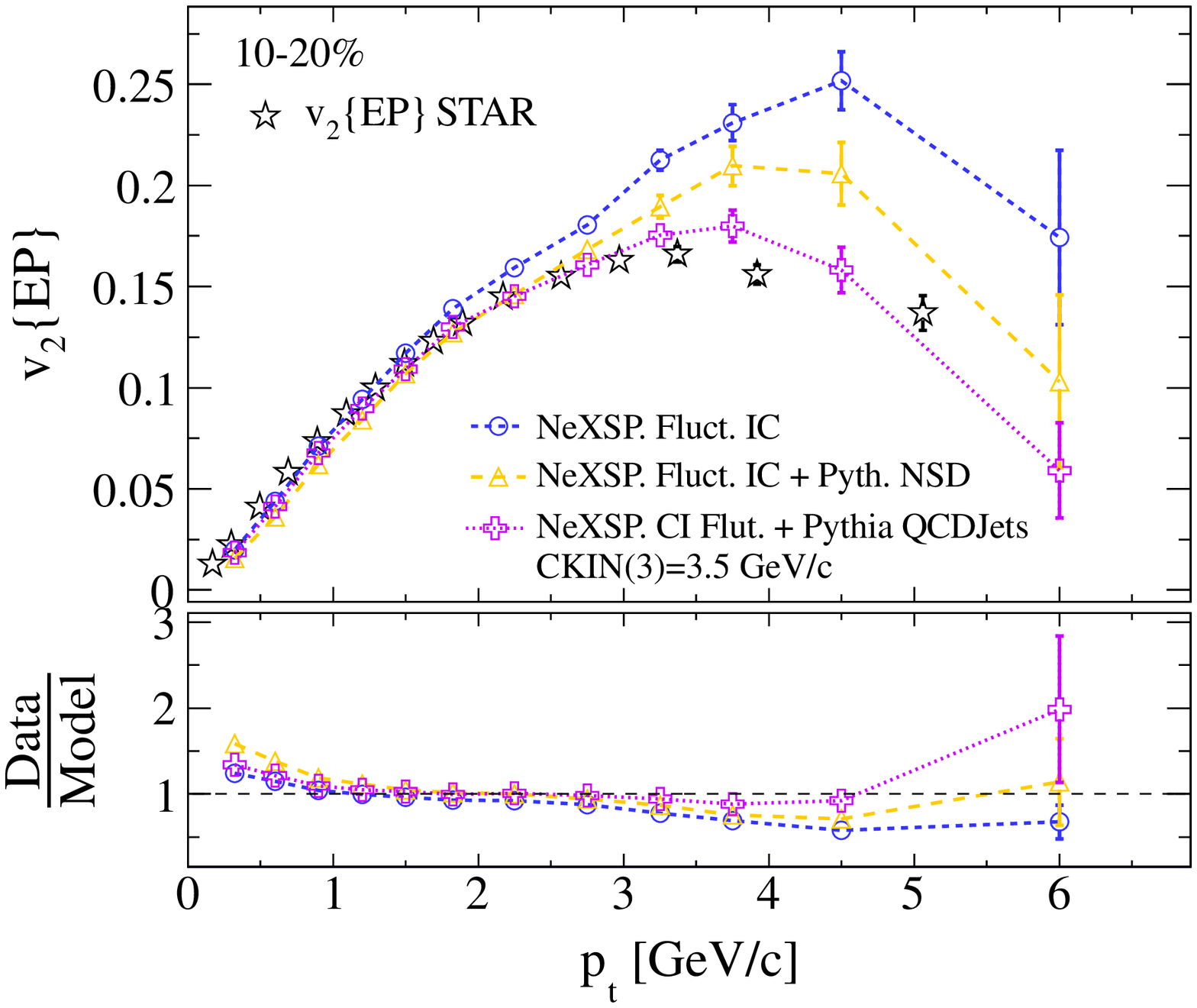}
\caption{\label{fig:v2EP_vs_pt}Differential $v_2$ from event plane method as a function of $p_t$ obtained with NeXSPheRIO+Pythia model compared to experimental results from STAR \cite{Adams:2004bi}.}
\end{minipage} 
\end{figure}

In order to construct a simplified two component model, we combine simulated events produced with NeXSPheRIO and Pythia event generators.
The NeXSPheRIO code \cite{Hama:2004rr,PhysRevC.65.054902} is based on ideal hydrodynamics and allows for the simulation of nucleus-nucleus collisions, while Pythia is an event generator for proton-proton (pp) collisions that takes into account the leading order processes in perturbative QCD \cite{sjostrand-2006-0605}.
We generated events of Au+Au collisions at $\sqrt{s_{NN}}=200$ GeV using the NeXSPheRIO code and embedded into each event an integer number $\alpha$ of p+p collisions generated using the Pythia code.
The number of Pythia events to be included was chosen so that the resulting sum of the Pythia and NeXSPheRIO charged pion spectra according to $p_t$ matched the experimentally measured $\pi^{\pm}$ spectra from STAR \cite{Abelev:2006jr}.
Also, we tested different Pythia configurations, namely NSD (non-single diffractive events) and QCD Jets (which prioritizes 2$\rightarrow$2 processes), as well as different cuts on the minimum transverse momentum of the main scattering process, referred here by CKIN(3) (see reference \cite{sjostrand-2006-0605} for details).
Figure \ref{fig:InvYield_vs_pt} shows the results for the 10-20\% centrality class for the various event generator configurations used.
Experimental points from STAR \cite{Abelev:2006jr} are shown for comparison.
The Monte Carlo configuration that best describes the spectra of charged pions was given by Pythia QCDJets with CKIN(3) = 3.5 GeV/c, for which $\alpha$ ranged from $\sim$1 to $\sim$9, being higher for more central collisions.
It is important to note that this procedure does not take into account any type of interaction of the jet with the medium.
The model still overestimates the intermediate $p_t$ region ($\sim 5$ GeV/c) and underestimates the very high $p_t$ region.
This behavior is consistent with the fact that we do not include energy loss of the jet particles to the medium, which would transfer a fraction of the momentum of these very energetic particles to lower momentum particles, thus populating the intermediate region and depleting the high $p_t$ spectrum.

\section{Effects on Flow Observables}
Using a sample of simulated events generated with the NeXSPheRIO+Pythia combination, we calculated the elliptic flow of the final particles produced.
The results are shown in figure \ref{fig:v2EP_vs_pt} as a function of $p_t$ for the centrality class 10-20\%.
Experimental points from STAR \cite{Adams:2004bi} are also shown for comparison.
As can be seen, pure NeXSPheRIO particles are not able to describe the experimental results for $p_t$ above 2.0 GeV/c.
The inclusion of Pythia events with large transferred momentum decreases the values of $v_2$ for high $p_t$.

\begin{figure}[h]
\begin{minipage}{17pc}
\includegraphics[width=15pc]{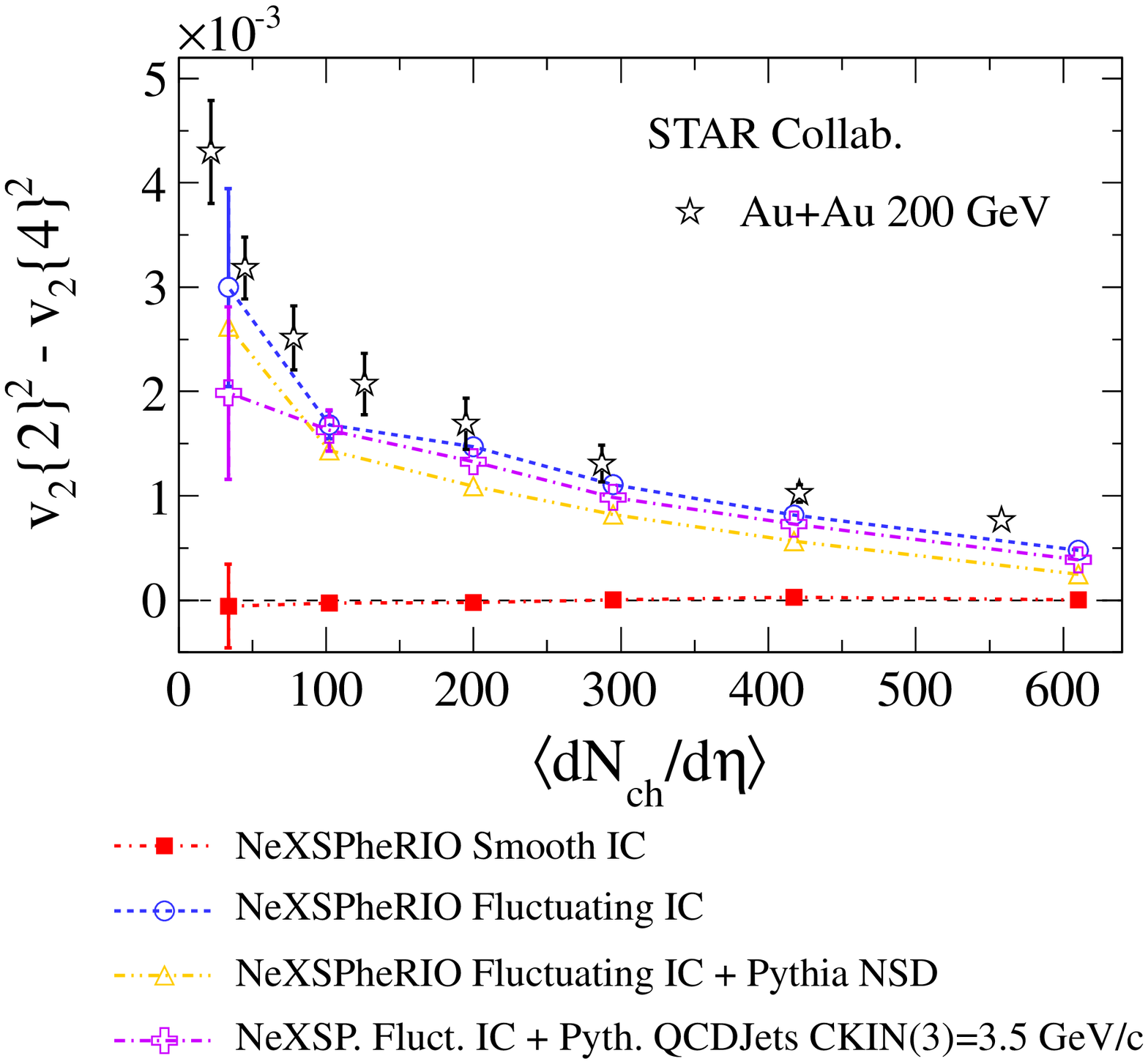}
\caption{\label{fig:v2Fluct_vs_dNchdeta}Difference between $v_2\{2\}^2$ and $v_2\{4\}^2$ as a function of centrality for different configurations of NeXSPheRIO+Pythia model. Experimental results from STAR \cite{Agakishiev:2011eq} are shown for comparison.}
\end{minipage}\hspace{2pc}
\begin{minipage}{17pc}
\includegraphics[width=15pc]{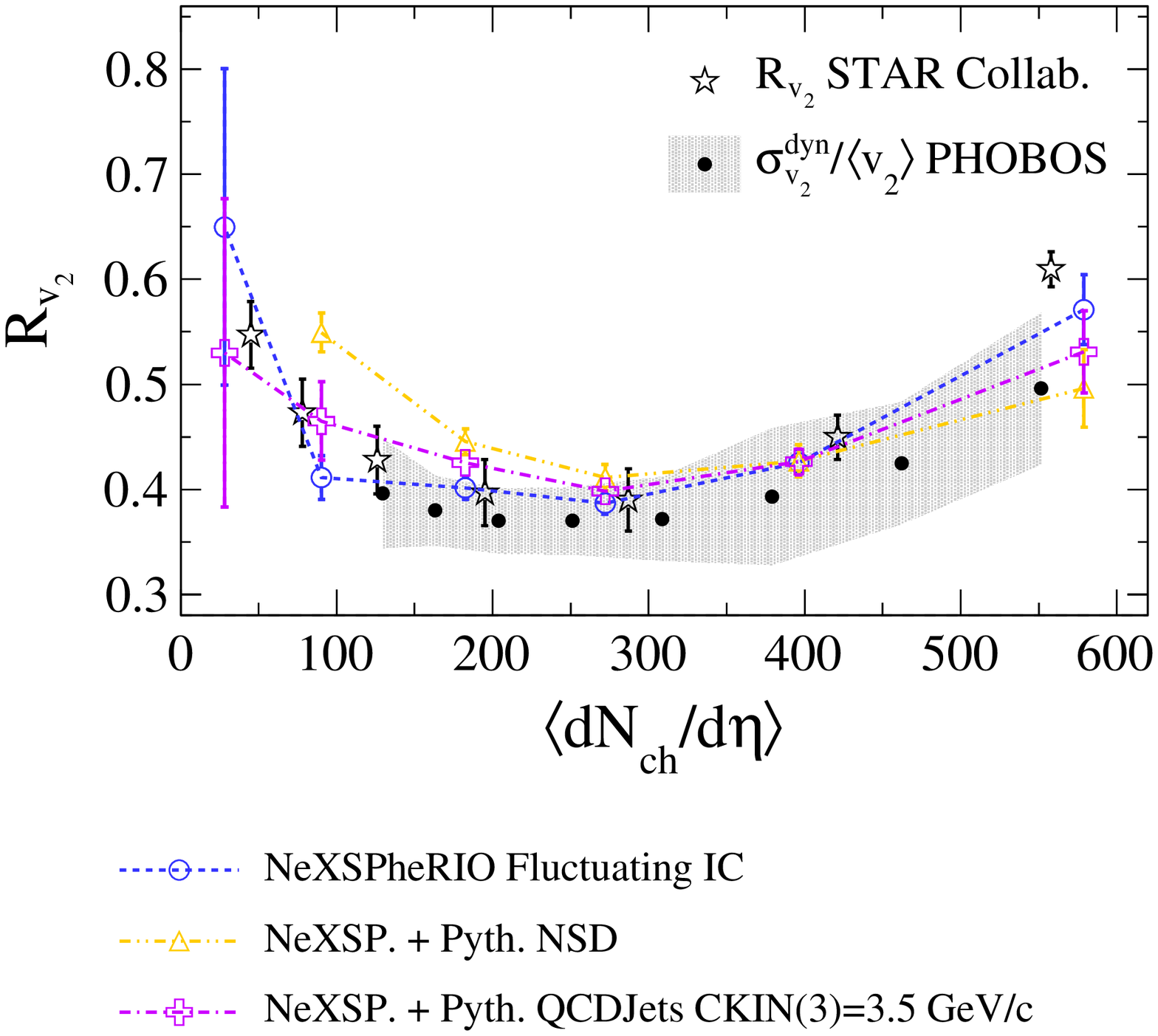}
\caption{\label{fig:Rv_vs_dNchdeta}$R_v$ parameter as a function of centrality for different configurations of NeXSPheRIO+Pythia model. Experimental results from STAR \cite{Agakishiev:2011eq} and PHOBOS \cite{Alver:2007qw} are shown for comparisons.}
\end{minipage} 
\end{figure}

In order to explore the effects produced by the superposition of NeXSPheRIO and Pythia events, we also calculated the difference between the squares of $v_2$ estimates of second and fourth orders obtained with the cumulant method, namely $v_2\{2\}^2-v_2\{4\}^2$.
This quantity has been shown to be proportional to the flow fluctuation ($\sigma_{v_2}$) and the non-flow contribution ($\delta$) \cite{Ollitrault:2009ie}, the latter being mainly dominated by jets.
The results obtained as a function of average number of charged particles produced at $\eta=0$ are shown in figure \ref{fig:v2Fluct_vs_dNchdeta}.
To illustrate the sensitivity of $v_2\{2\}^2-v_2\{4\}^2$ to the fluctuations and non-flow contribution we also included the results obtained for NeXSPheRIO events generated with smoothed initial conditions, which produce values consistent with zero.
However, contrary to what we expected, the addition of Pythia events did not increase the values of these parameters, suggesting that the non-flow contribution may be negligible for the integrated flow estimates.
In figure \ref{fig:Rv_vs_dNchdeta} we show the results obtained for the $R_{v_2}$ parameter, defined as the square-root of $(v_2\{2\}^2-v_2\{4\}^2)/(v_2\{2\}^2+v_2\{4\}^2)$.
Again, the inclusion of Pythia particles in the NeXSPheRIO events does not seem to produce a relevant change in the behavior of $R_{v_2}$.
For comparison, experimental results from STAR \cite{Agakishiev:2011eq} and PHOBOS \cite{Alver:2007qw} are also shown.

\section{2-Particle Correlation}
\begin{figure}[h]
\begin{minipage}{24pc}
\includegraphics[width=12pc]{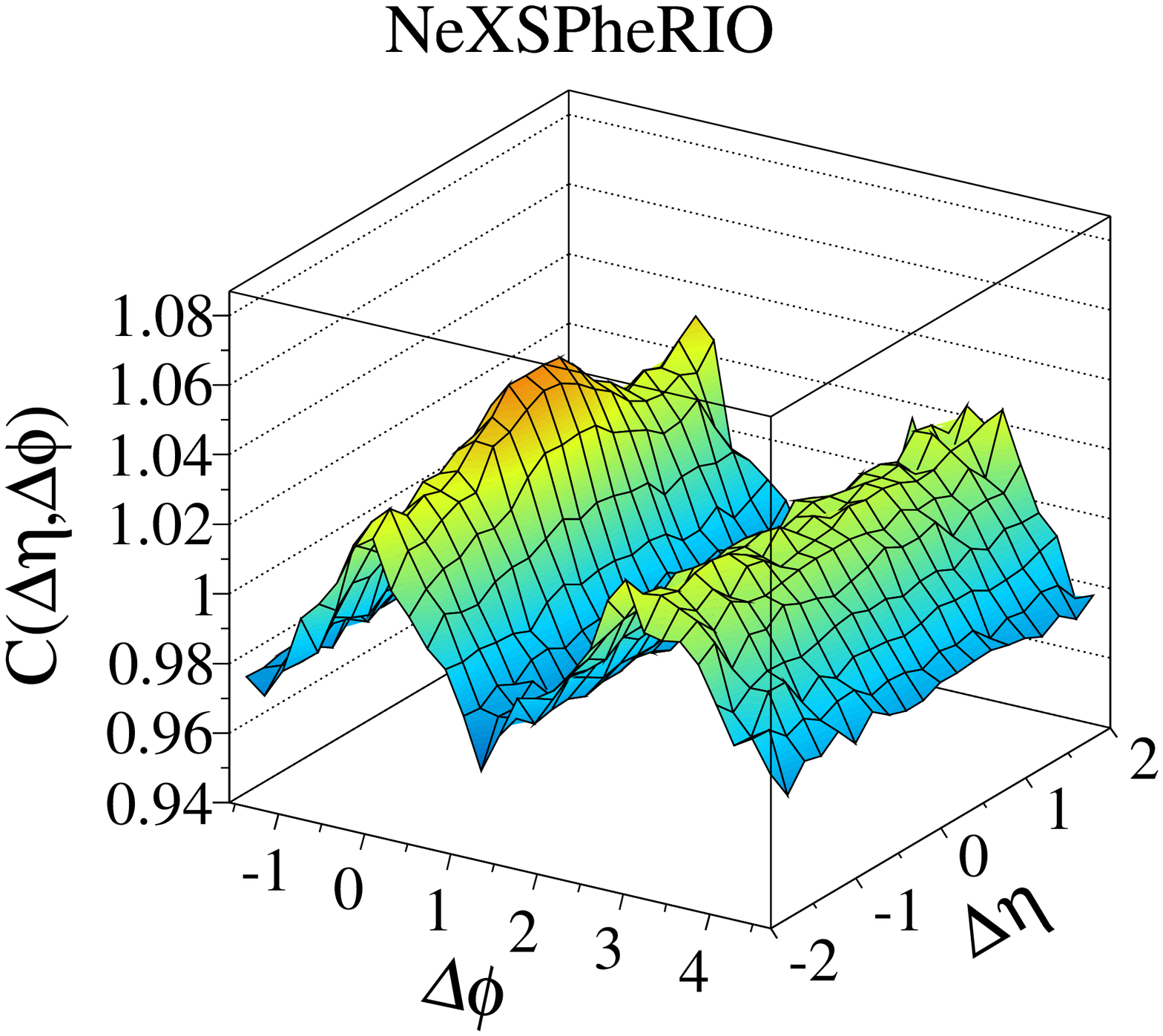}
\includegraphics[width=12pc]{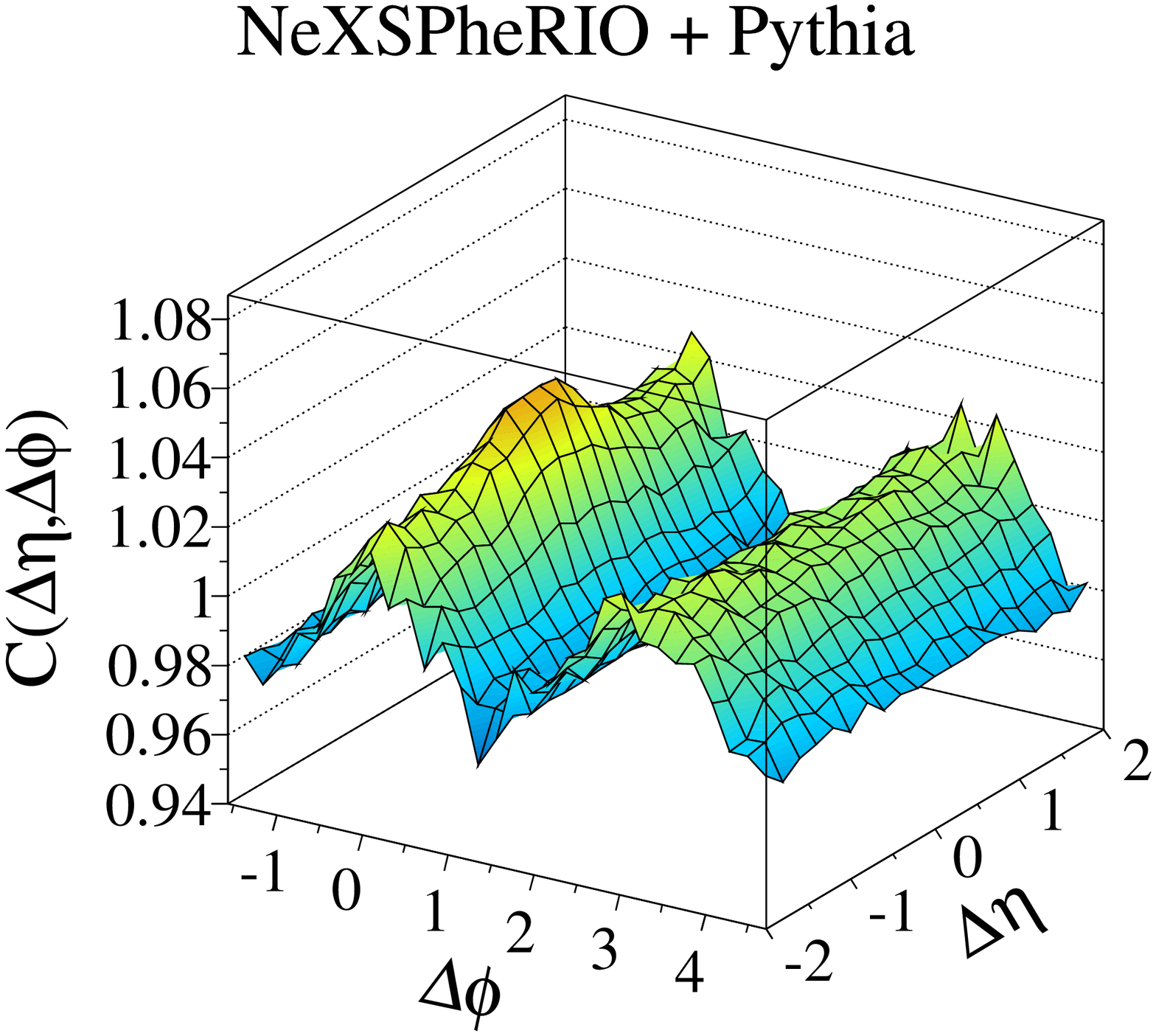}
\caption{\label{fig:CorrC_vs_DEtaDPhi}Two-particle correlation distribuition obtained from NeXSPheRIO with fluctuating initial condition (left plot) and from NeXSPheRIO+Pythia for QCDJets with CKIN(3)=3.5 GeV/c configuration (right plot).}
\end{minipage}\hspace{0.7pc}
\begin{minipage}{13pc}
\includegraphics[width=12pc]{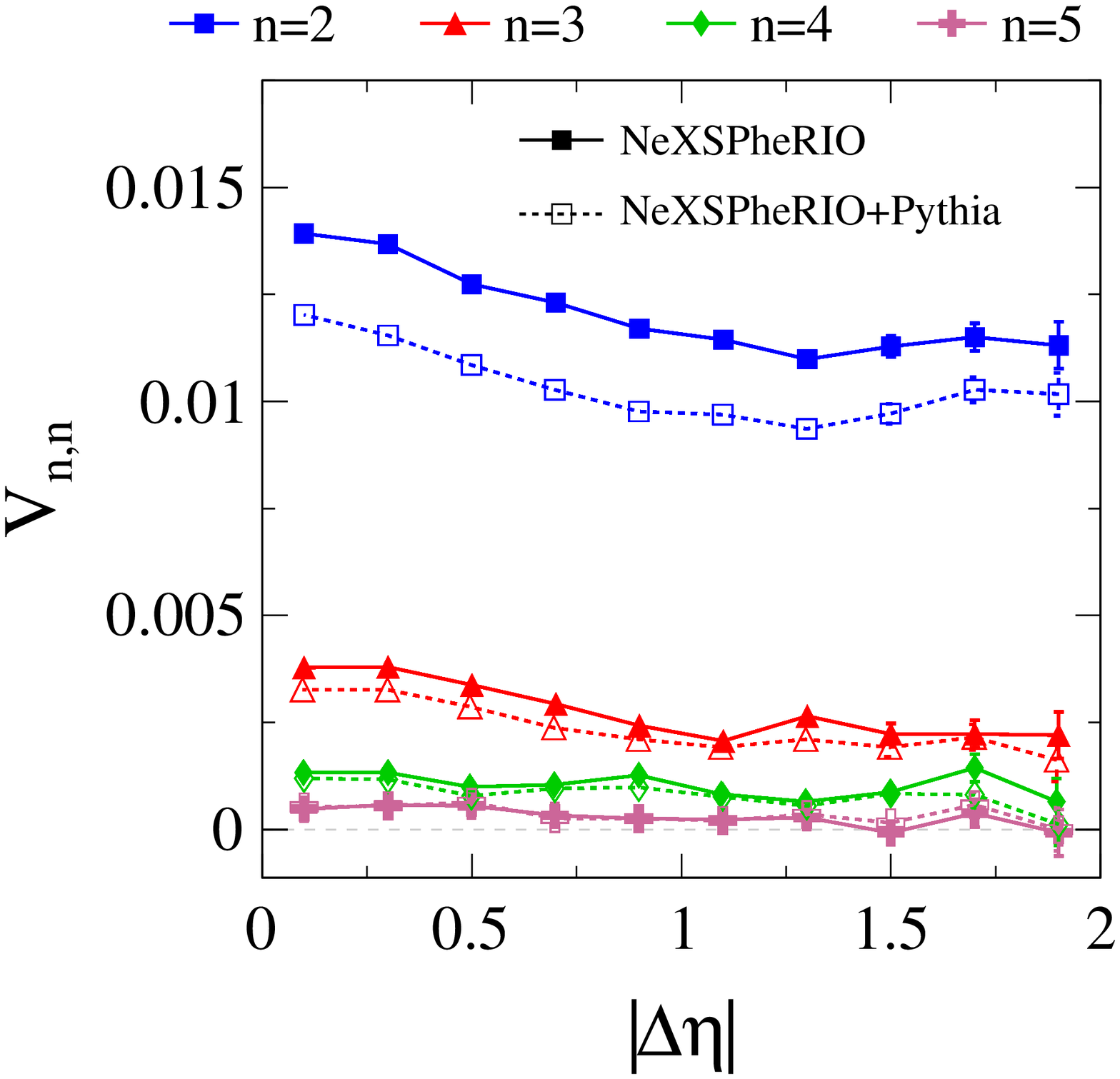}
\caption{\label{fig:Vnn_vs_DEta}Results of fits over slices of the two-particle distributions for NeXSPheRIO and NeXSPheRIO+Pythia.}
\end{minipage} 
\end{figure}

Another interesting result from particle correlations in heavy ion collision experiments is the observation of the ridge \cite{Putschke2007507,McCumber:2008id}.
Initially thought to be an effect of jets traversing the medium created in the collision, it was shown later that it could also be described within a hydrodynamic treatment considering fluctuating initial conditions \cite{Takahashi:2009na}.
However, both mechanisms may be present, and thus it is of interest to look into how the inclusion of Pythia affects the shape of the $\Delta\eta\Delta\phi$ distribuition.
We define a two-particle correlation function in the same way as presented in reference \cite{ATLAS:2012at}, $C(\Delta\eta,\Delta\phi)=S(\Delta\eta,\Delta\phi)/B(\Delta\eta,\Delta\phi)$,
where $S$ refers to the same-event pair distribution and $B$ is the combinatorial pair distribution obtained by mixing particles $a$ and $b$ from different events.
The results obtained using pure NeXSPheRIO events (left plot) and NeXSPheRIO+Pythia events (right plot) are shown in figure \ref{fig:CorrC_vs_DEtaDPhi}.
The correlation was calculated using charged particles with $0.5 < p_t^a < 1.5$ GeV/c (associated) and $p_t^b > 2.0$ GeV/c (trigger).
In order to quantify the effects of including decorrelated Pythia events we performed a fit over slices of $\Delta\eta$ using the sum of the first five harmonics in $dN_{pairs}/d\Delta\phi \propto 1 + 2 \sum V_{n,n} \cos(n\Delta\phi)$.
The results are shown in figure \ref{fig:Vnn_vs_DEta} for pure NeXSPheRIO and for NeXSPheRIO+Pythia.
Although Pythia particles are highly correlated within each Pythia event, the incoherent addition of a number of Pythia events to a NeXSPheRIO event drives the coefficients of the harmonic decomposition down.

\section{Summary and Conclusions}
In this work, we have investigated a model in which a number of Pythia events is added to a single NeXSPheRIO Au+Au event and have shown that, by using a number of Pythia events so that the resulting spectra match experimental data, a better description of flow variables is also obtained.
In addition, to quantify the non-flow effects over particle correlation analysis, we evaluated the difference $v_2\{2\}^2-v_2\{4\}^2$, for integrated flow as a function of centrality as well as the harmonic decomposition of two-particle correlation distribution for a specific choice of $p_t$ range for associated and trigger particles.
The results obtained suggest that the effects over integrated quantities are very small and the observed behavior is mainly due to fluctuations in the initial condition.
On the other hand, when we focus on the intermediate to high $p_t$ region, the effects start to appear.
However, it actually lowers the correlation observed, indicating that the fact that the extra added particles are added in an uncorrelated way has a stronger effect on the observables than the correlation among the few added particles.
In conclusion, it is interesting to note that experimental results of $v_2$ are better described for higher values of $p_t$ by adding independent jet-like proton-proton collisions. 
It is clear that a more detailed investigation using a similar approach may be important to understand experimental observations.

\ack
This work was supported in part by FAPESP under Project No. 2012/02895-2 and by CNPq of Brazil.
The authors wish to thank Dr. Takeshi Kodama for fruitful discussions.

\section*{References}
\bibliography{procSQM2013.bib}

\end{document}